# Magnetocardiography measurements using an optically pumped magnetometer under ambient conditions

Kushal Patel[1], Kesavaraja C[1], Pranab Dutta[1], Korak Biswas[1]

## Abstract

In this work, we report the development of a rubidium-based single-beam scalar optically pumped magnetometer (OPM) and demonstrate its application in measuring human cardiac magnetic fields in an unshielded environment. The developed magnetometers exhibit a noise floor below 15 pT/$\sqrt{\text{Hz}}$ in the frequency range of 1–35 Hz, with a measurement bandwidth of 100 Hz. When operated in a gradiometric configuration, the noise floor is further reduced to below 3 pT/$\sqrt{\text{Hz}}$ over the same frequency range. Magnetocardiography (MCG) signals were recorded at five different locations across the thorax. A clear polarity reversal of the QRS complex was observed across these measurement positions, confirming the spatial sensitivity of the system. The proposed system shows strong potential for clinical diagnostics, offering valuable physiological information through non-contact MCG measurements

## Introduction

Magnetocardiography (MCG) is a non-contact and non-invasive technique used to record the magnetic fields generated by the electrical activity of the heart. Unlike electrocardiography (ECG), which records the potential difference between electrodes placed on the body surface, MCG detects the associated magnetic fields without requiring direct electrical contact [1]. Owing to the advantage of MCG signals are minimally influenced by the conductive properties of biological tissues, MCG has several advantages particularly in localizing cardiac current sources and identifying abnormalities such as ischemia and coronary artery disease, as reported in the literature. Despite these advantages, MCG has not yet been adopted for routine clinical practice, primarily due to the high cost of ultra-sensitive magnetic sensors and the requirement of magnetically shielded rooms to detect the extremely weak cardiac magnetic fields [2, 3].

The development of MCG began in the late 1980s with the advent of low-temperature superconducting quantum interference device (SQUID) sensors, which enabled the detection of cardiac magnetic signals but required cryogenic cooling for operation [4,5].

Room-temperature magnetic sensors such as tunnel magnetoresistance, fluxgate, Nitrogen - vacancy centres in diamond and induction-coil sensors have also been explored for MCG applications, offering sensitivities in the picotesla (pT) range [6-10]. However, these sensors are still in early stages of development and their performance in unshielded environments remains strongly affected by environmental magnetic noise.

In recent years, atomic magnetometers based on alkali-metal vapors have emerged as promising alternatives for biomagnetic signal measurements. Specifically, Spin-exchange relaxation-free (SERF) optically pumped magnetometers (OPMs) have demonstrated sensitivities comparable to SQUID systems, but typically require near-zero magnetic field conditions [11]. In addition to that OPMs operating in unshielded environments such as Mx, Mz, Bell-Bloom and scalar free induction decay (FID) OPMs are also explored [12]. Among these, scalar OPMs offer a comparatively simple configuration and the capability to operate in unshielded environments while maintaining high sensitivity [13]. Furthermore, when operated in a gradiometric configuration, scalar OPMs can effectively suppress common-mode environmental noise, including fluctuations in the Earth's magnetic field without substantially increasing hardware complexity, making it attractive for practical and potentially portable MCG systems [14 - 16].

The present study focuses on the development of a scalar OPM in gradiometric arrangement and its performance in unshielded environment. The proposed sensor system has been validated experimentally for detecting human MCG signals in an unshielded environment.

## Materials and Methods

### Experimental setup

MCG measurements were carried out using an in-house developed rubidium-based scalar OPM system. A single laser beam was fiber-coupled and split to illuminate two rubidium vapor cells, which were separated by a distance of 35 mm and operated as a gradiometric sensor pair as in the Figure 1(a). The same laser beam served as both the optical pump and probe for atomic spin polarization and signal detection. The rubidium atoms within each    vapor cell were thermally vaporized using laser-based heating, with the cells enclosed by graphene-coated layers to ensure uniform temperature distribution and stable operation. This configuration enabled room-temperature operation of the magnetometers while maintaining high sensitivity. Two identical magnetometers were operated in a gradiometric configuration to suppress environmental magnetic noise during unshielded measurements. The primary sensor was positioned close to the subject's chest to detect cardiac magnetic fields, while the secondary sensor acted as a reference channel to capture common-mode background magnetic fluctuations. Subtraction of the two sensor outputs yielded the magnetic field gradient, effectively reducing ambient noise contributions as seen in the Figure 1(b).

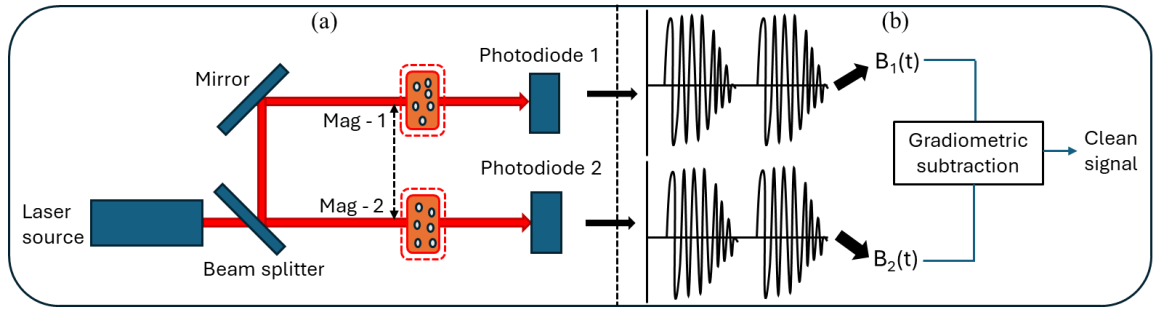

Figure 1: (a) Schematic of scalar OPM setup (b) Post-processing of OPM signals.

## Data acquisition

The vapor cell output of the scalar OPM was recorded as signals using Pico scope with sampling rate of 5 MHz. These signals were processed by fitting a nonlinear exponential decay model to extract the Larmor precession frequency. The time-varying Larmor frequency was then converted into a magnetic field time series, representing the measured magnetic fields with a bandwidth of 100 Hz as in Figure 1(b).

The subject was seated comfortably on a non-magnetic plastic chair to avoid magnetic interference. Five predefined measurement locations were marked on the thoracic region as in the Figure 2(a). At each location, the primary sensor was aligned as close as possible to the chest surface, resulting in a sensor–skin standoff distance of less than 1 cm as displayed in Figure 2(b). The sensor position was maintained constant during each recording to minimize motion-related artifacts. The single-lead ECG signals were recorded simultaneously along with the MCG signals to reliably identify cardiac cycles and improve signal quality. The ECG was acquired in Lead I configuration using an Arduino-based data acquisition system. Both ECG and MCG signals were recorded synchronously throughout the measurement period of five minutes for each measurement location. All measurements were carried out on a healthy adult volunteer (Age: 35, Male) with no reported history of cardiovascular disease. The subject exhibited normal sinus rhythm throughout the measurement time. The data analysis was performed using custom-built programs developed in Python.

## Signal processing

The signal processing steps used to obtain the final MCG signal are summarized in Figure 3. The approach combines magnetic field measurements with simultaneously recorded ECG signals. The gradiometric MCG time series and the ECG time series were first pre-processed to remove unwanted frequency components. Both signals were digitally filtered with 50 Hz notch filter and bandpass filtered in the 1–40 Hz range with Finite impulse response filter with 201 taps, which preserves the dominant cardiac features while reducing baseline drift and high-frequency noise.

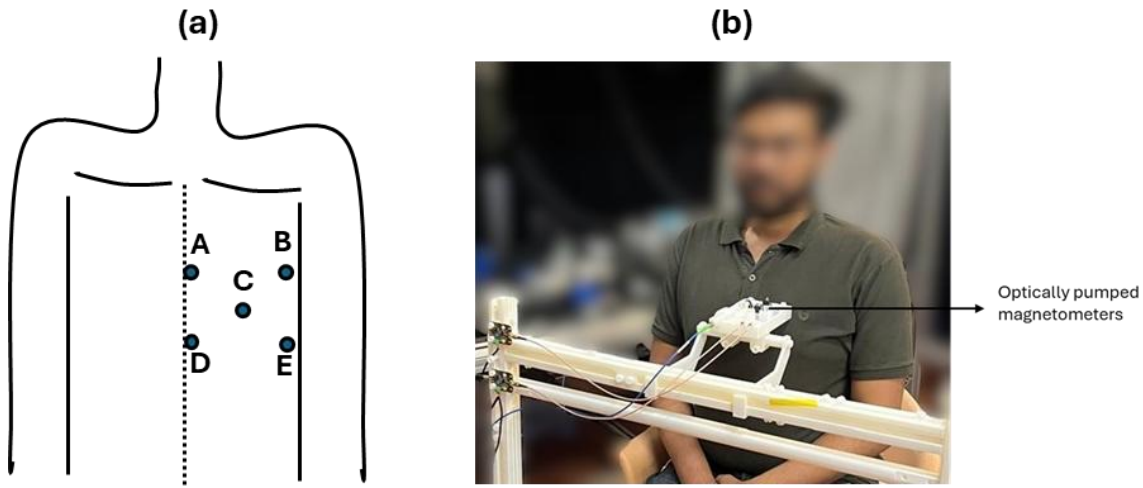

Figure 2: (a) MCG recording positions on the thorax and (b) A subject sitting close to the OPM sensors for measurement position A.

Cardiac cycles were identified using the ECG signal. R-peaks were detected from the filtered ECG using a standard Pan Tomkins algorithm [17], and these time points were used as reference markers to segment the corresponding MCG signal into individual cardiac cycles. This ensured accurate temporal alignment between the electrical and magnetic signatures of cardiac activity. The segmented MCG cycles were then combined using synchronized averaging. Since the cardiac magnetic signal is repeatable from beat to beat, this averaging process enhances consistent physiological components while suppressing random noise. To further improve robustness, beats which were having correlation threshold (>0.3) with the averaged cardiac cycle artifacts were identified and excluded from the analysis. Finally, the remaining clean cycles were averaged to obtain the signal-averaged MCG waveform. This averaged signal provides a clearer representation of the underlying cardiac magnetic activity and is suitable for subsequent analysis and interpretation.

## Results and Discussion

The noise performance of the developed scalar optically pumped magnetometer system was first evaluated. The noise floor densities of the two individual magnetometers were measured to be 14.26 pT/$\sqrt{\text{Hz}}$ and 13.98 pT/$\sqrt{\text{Hz}}$, respectively, within the frequency range of 1–35 Hz. When operated in a gradiometric configuration, common background magnetic noise was effectively suppressed, resulting in a common-mode rejection ratio (CMRR) of 32.34 dB. This gradiometric subtraction reduced the effective noise floor to 2.61 pT/$\sqrt{\text{Hz}}$ over the same frequency range, as shown in Figure 4.

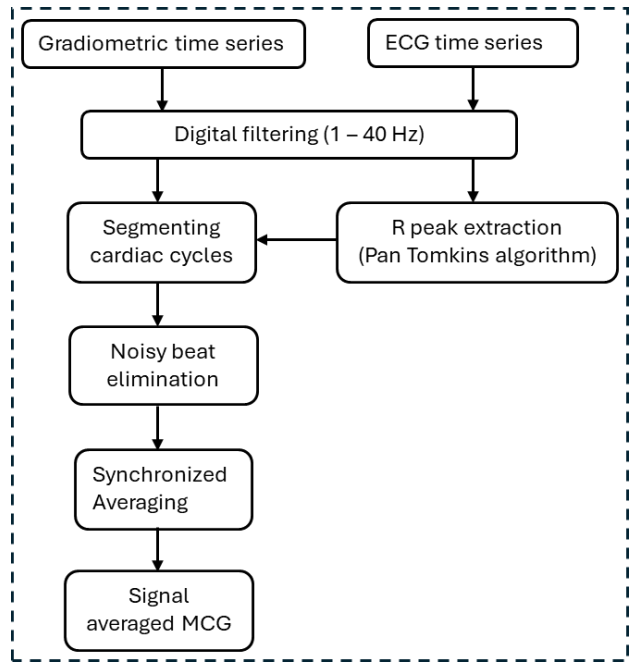

Figure 3: Flow chart of the methodology followed in the present study.

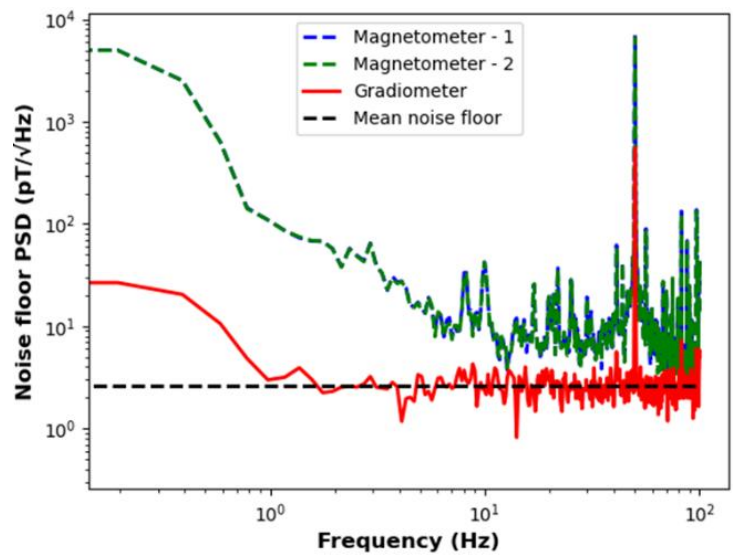

Figure 4: Noise floor density of the magnetometers and their gradiometric output measured in an unshielded environment.

In addition to PSD estimation, the noise present in the environment measured by the system at different time scales is further explored using Allan deviation curves as in Figure 5. At short averaging times, the Allan deviation decreases with increasing τ for all three cases, indicating that white noise is the dominant contribution in this region and that temporal averaging effectively reduces random fluctuations.

As the averaging time increases further, the Allan deviation of the individual magnetometers begins to rise, reflecting the influence of low-frequency noise components such as flicker noise and drift. In contrast, the gradiometric output maintains a comparatively lower deviation and shows improved stability over a broader range of averaging times.

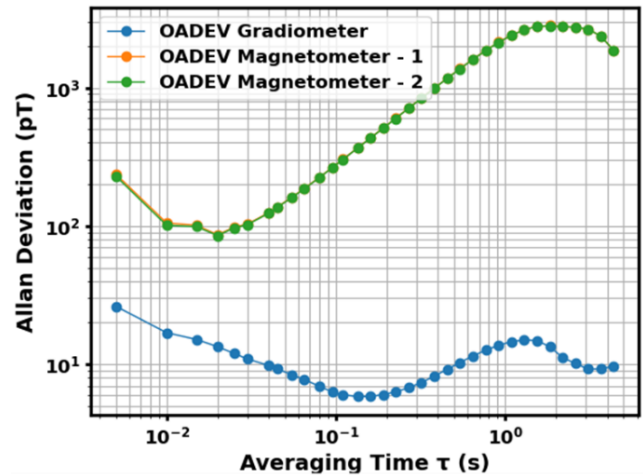

Figure 5: Allan deviation of the two individual magnetometer and the gradiometric output.

The gradiometer balance and the performance of the noise cancellation offered by such configuration are also evaluated in the Allan deviation curves of balance ratio and the CMRR. In Figure 6(a), the balance ratio decreases with increasing averaging time, particularly in the short-to-intermediate τ region. This trend indicates that temporal averaging improves the matching between the two magnetometer channels by reducing random noise contributions. At longer averaging times, the balance ratio approaches a stable low value, suggesting that the residual mismatch is primarily governed by slow drift components rather than random fluctuations.

Similarly, Figure 6(b) shows the variation of CMRR with averaging time. The CMRR increases progressively with τ, reaching its maximum at intermediate averaging times. This improvement reflects enhanced suppression of common-mode environmental noise as uncorrelated noise components are reduced through averaging. At longer averaging times, the CMRR tends to saturate, indicating that further averaging does not significantly enhance rejection performance due to underlying low-frequency drift and sensor mismatch. The combined analysis of balance ratio and CMRR demonstrates the effectiveness of the gradiometric configuration in improving environmental noise suppression and overall system stability.

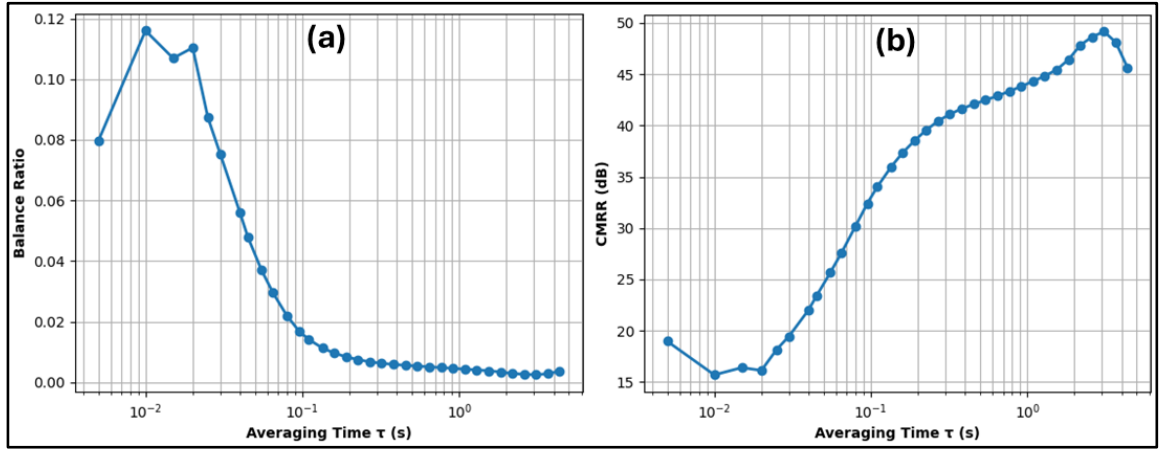

Figure 6: Allan deviation of the gradiometer balance and the CMRR for different averaging time.

The pre-processed signal averaged MCG signals measured at five different locations are displayed in Figure 7. Clear cardiac features, including the QRS complex, are visible in the average signals, with noticeable variations in amplitude and polarity across measurement locations.

In order to improve visualization of finer waveform features, particularly around the QRS complex and T-wave regions, a Savitsky–Golay filter was applied to smooth residual high-frequency fluctuations [18]. The smoothed MCG signals are overlaid and compared with the corresponding average ECG waveform in Figure 8 (a). The magnetic field map was generated at the R-peak time instant by spatially interpolating the measured magnetic field values across the sensor array, resulting in a continuous contour representation with a uniform colour scale as displayed in Figure 8 (b).

The signal quality was further quantified by estimating the signal-to-noise ratio (SNR) of the QRS complex. The noise level was defined using the baseline segment preceding the P-wave, while the QRS peak-to-peak amplitude was taken as the signal component. The noise level was defined using the baseline segment preceding the P-wave, while the QRS peak-to-peak amplitude was taken as the signal component. The SNR improved with an increasing number of averaged beats and stabilized at a mean value of approximately 10 dB as in Figure 9. This behaviour confirms the benefit of synchronized averaging for enhancing weak cardiac magnetic signals under unshielded measurement conditions.

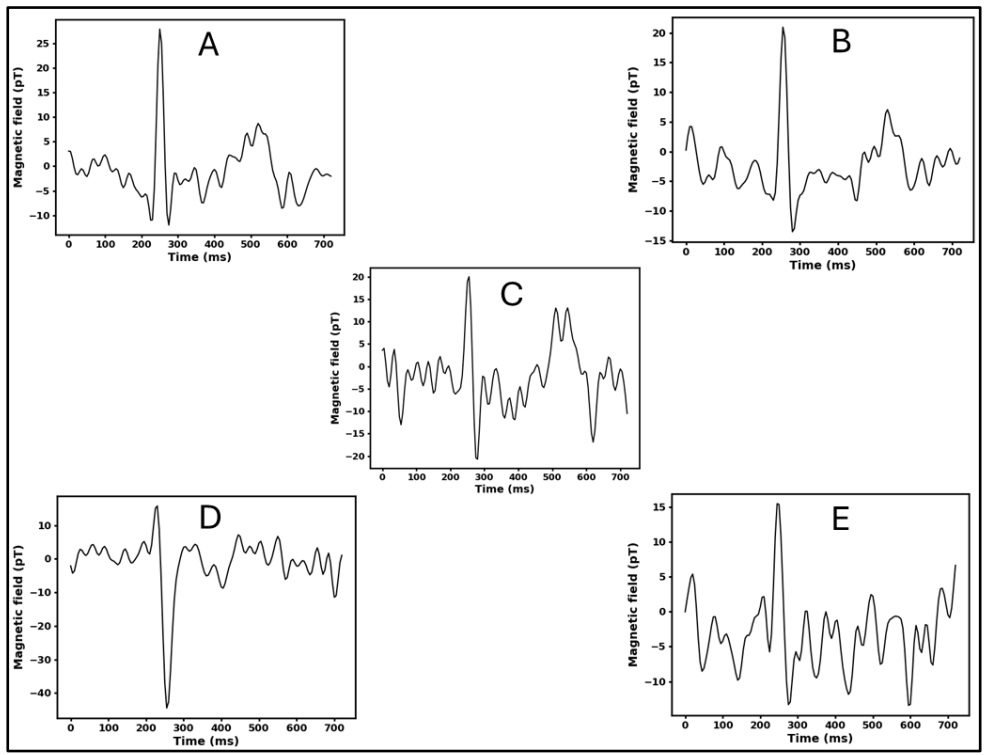

Figure 7: The signal averaged MCG signals measured at five different locations A, B, C, D and E on the thorax.

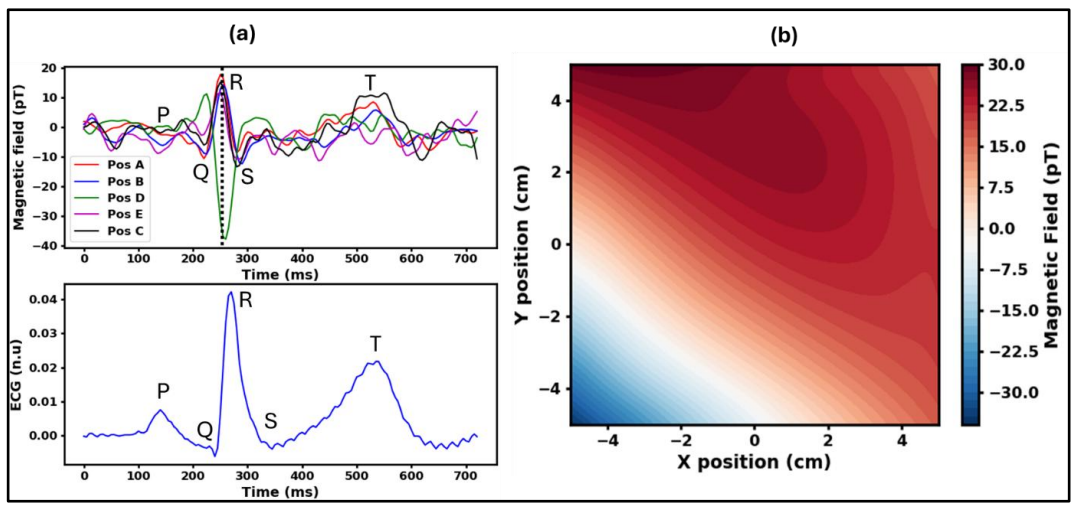

Figure 8: (a) The smoothed version of the signals using Savitsky - Golay filter is overlay plotted for better visualization against the corresponding averaged ECG (b) The Spatio-temporal distribution of magnetic fields are plotted as a contour map taken at R peak time instant and also the pseudo current density vectors are plotted to show the direction of the current flow.

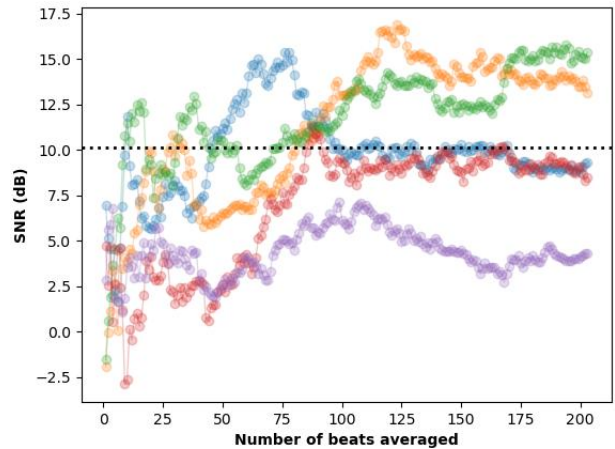

Figure 9: SNR of the QRS complex vs Number of beats averaged computed for MCG signals measured at different locations.

## Conclusion

This study demonstrates the feasibility of recording human MCG signals using a rubidium-based scalar OPM operated in a gradiometric configuration under unshielded conditions. The system achieved noise floor sensitivity of 2.6 pT/$\sqrt{\text{Hz}}$ through gradiometric noise suppression, enabling reliable detection of cardiac magnetic signals at multiple thoracic locations. These results indicate that simple, room-temperature scalar OPMs can provide sufficient sensitivity for non-contact cardiac magnetic measurements and offer a promising pathway toward compact and clinically accessible MCG systems. The study can be extended to improve the signal quality of other cardiac features such as T wave, P wave by increasing the number of channels with use of advanced signal processing methods [15].

## Acknowledgement

This research was supported by GDQLABS Private Limited. The authors gratefully acknowledge Dr. Megha Agarwal, CMTI, Bengaluru, India, and Prof. Umakant Rapol, IISER Pune, India, for their guidance, support, and helpful discussions.

## References

Malmivuo, Jaakko, and Robert Plonsey. "Bioelectromagnetism: principles and applications of bioelectric and biomagnetic fields". Oxford university press, 1995.

Fenici R, Brisinda D, Meloni AM. "Clinical application of magnetocardiography". Expert review of molecular diagnostics. 2005 May 1;5(3):291-313.

Li J, Shen Y, Shen C, et al. "Advances of magnetocardiography in application of adult and fetal cardiac diseases". Frontiers in Cardiovascular Medicine. 2025 Jul 16; 12:1522467.

Clarke J, Braginski AI, editors. "The SQUID handbook: Applications of SQUIDs and SQUID systems". John Wiley & Sons; 2006 Dec 13.

Sternickel K, Braginski AI. "Biomagnetism using SQUIDs: status and perspectives". Superconductor Science and Technology. 2006 Feb 15;19(3): S160.

Mooney JW, Ghasemi-Roudsari S, et al. "A portable diagnostic device for cardiac magnetic field mapping". Biomedical Physics & Engineering Express. 2017 Jan 13;3(1):015008.

Xing Z, Li H, Dou H, et al. "Multi-level Gated U-Net for Denoising TMR Sensor-Based MCG Signals". International Conference on Medical Image Computing and Computer-Assisted Intervention 2025 Sep 21 (pp. 428-437). Cham: Springer Nature Switzerland.

Reermann J, Elzenheimer E, et al. "Real-time biomagnetic signal processing for uncooled magnetometers in cardiology". IEEE Sensors Journal. 2019 Jan 17;19(11):4237-49.

Omar M, Benke M, et al. "Human Cardiac Measurements with Diamond Magnetometers". arXiv preprint arXiv:2601.18843. 2026 Jan 26.

Yaga L, Amemiya M, et al. "Recording of Cardiac Excitation Using a Novel Magnetocardiography System with Magnetoresistive Sensors" Outside a Magnetic Shielded Room. Sensors. 2025 Jul 26;25(15):4642.

Shah VK, Wakai RT. "A compact, high performance atomic magnetometer for biomedical applications". Physics in Medicine & Biology. 2013 Nov 7;58(22):8153.

Fabricant A, Novikova I, Bison G. "How to build a magnetometer with thermal atomic vapor: a tutorial". New Journal of Physics. 2023 Feb 20;25(2):025001.

Clancy RJ, Gerginov V, et al. "A study of scalar optically-pumped magnetometers for use in magnetoencephalography without shielding". Physics in Medicine & Biology. 2021 Sep 3;66(17):175030.

Limes ME, Foley EL, et al. "Portable magnetometry for detection of biomagnetism in ambient environments". Physical Review Applied. 2020 Jul 1;14(1):011002.

Iwata, Geoffrey Z., et al. "Bedside Magnetocardiography with a Scalar Sensor Array." Sensors 24.16 (2024): 5402.

Xiao W, Sun C, et al. "A movable unshielded magnetocardiography system". Science Advances. 2023 Mar 29;9(13): eadg1746.

Pan J, Tompkins WJ. "A real-time QRS detection algorithm". IEEE transactions on biomedical engineering. 2007 Mar 12(3):230-6.

Schafer RW. "What is a savitzky-golay filter?" [lecture notes]. IEEE Signal processing magazine. 2011 Jun 16;28(4):111-7.